%% file: manuscript.tex
\documentclass[apjl]{emulateapj}
\usepackage{comment}
\usepackage{ifthen}
\usepackage{amsmath}
\usepackage{amsfonts}
\usepackage{amssymb}
\usepackage{multirow}
\usepackage{wasysym}
\usepackage{subfigure}
\usepackage{epsfig}

\input{shortcuts.tex}

\newboolean{emulateapj}
\setboolean{emulateapj}{true}

\newboolean{astroph}
\setboolean{astroph}{true}

\newboolean{png}
\setboolean{png}{false}

\shortauthors{Sliski \& Kipping}
\shorttitle{High False Positive Rate for KOIs of Giants}
\ifthenelse{\boolean{emulateapj}}{
    \newcommand{\titledag}{$\dagger$}
}{
    \newcommand{\titledag}{\dagger}
}

\begin{document}

\title {A High False Positive Rate for \emph{Kepler} Planetary Candidates \\
of Giant Stars using Asterodensity Profiling
\altaffilmark{\titledag}}

\author{
	{\bf David H. Sliski$^{1,*}$ and David M. Kipping$^{1,2}$}
}
\altaffiltext{1}{Harvard-Smithsonian Center for Astrophysics, 60 Garden St, Cambridge, MA 02138, USA}
\altaffiltext{2}{Sagan Fellow}
\altaffiltext{*}{dsliski@cfa.harvard.edu}

\altaffiltext{$\dagger$}{
Based on archival data of the \emph{Kepler} telescope. 
}


\begin{abstract}

Asterodensity Profiling (AP) is a relatively new technique for studying transit
light curves. By comparing the mean stellar density derived from the transit
light curve to that found through some independent method, AP provides
information on several useful properties such as orbital eccentricity and 
blended light. We present an AP survey of 41 \emph{Kepler} Objects of Interest 
(KOIs), with a single transiting candidate, for which the target star's mean 
stellar density has been measured using asteroseismology. The ensemble 
distribution of the AP measurements for the 31 dwarf stars in our sample shows 
excellent agreement with the spread expected if the KOIs were genuine and have 
realistic eccentricities. In contrast, the same test for the 10 giants in our 
sample reveals significant incompatibility at $>4$\,$\sigma$ confidence. Whilst 
extreme eccentricities could be invoked, this hypothesis requires four of the 
KOIs to contact their host star at periastron passage, including the recently 
claimed confirmation of Kepler-91b. After carefully examining several 
hypotheses, we conclude that the most plausible explanation is that the 
transiting objects orbit a different star to that measured with asteroseismology 
- cases we define as false-positives. Based on the AP distribution, we estimate 
a false-positive rate (FPR) for \emph{Kepler's} giant stars with a single transiting 
object of FPR$\simeq70\%\pm30$\%.

\end{abstract}



\section{INTRODUCTION}
\label{sec:intro}

Over the last five years \emph{Kepler} has revolutionized our understanding of 
exoplanetary systems with the discovery of several thousand planetary 
candidates\footnote{See http://archive.stsci.edu/kepler/planet\_candidates.html}.
One of the revelations to have emerged from this avalanche of objects is
that conventional follow-up techniques for transit surveys, such as radial
velocity observations, are impractical and prohibitively expensive when faced
with several thousand targets. For this reason, great effort has been spent
to find alternative methods to validate planetary systems which can ideally
make use of the original \emph{Kepler} data alone, such as
blend analysis \citep{torres:2011}, Transit Timing Variations (TTV) 
\citep{nesvorny:2012}, planetary reflection/emission 
\citep{esteves:2013} and validation by multiplicity \citep{lissauer:2014,
rowe:2014}. With the upcoming TESS \citep{ricker:2010} and PLATO 
\citep{rauer:2013} missions expected to discover tens of thousands more 
planetary candidates, these techniques will be of great value to the wider 
community. In this vein, a new technique dubbed ``Asterodensity Profiling'' (AP) 
has recently been proposed as a tool to both aid in planetary validation and 
measuring the orbital eccentricity of planetary candidates (see 
\citealt{MAP:2012,dawson:2012,AP:2014} and references therein).

AP exploits the fact that one can infer the mean stellar density of a star,
$\rho_{\star,\obs}$, from the shape of a transit light curve under various 
idealized assumptions, as first demonstrated by \citet{seager:2003}. If one
has some independent measure of the mean density, $\rho_{\star,\tru}$, a direct
comparison allows one to test these idealized assumptions and ultimately
extract useful information about the properties of the system. For example, 
\citet{dawson:2012} showed how the orbital eccentricity of a planet may be
inferred using the so-called ``photo-eccentric'' effect and \citet{AP:2014}
showed how the quantity of blended light may be constrained using the
``photo-blend'' effect. Generally, cases where the two derived densities are 
dramatically different are the most interesting, since these immediately imply 
that the idealized assumptions cannot hold \citep{dawson:2013}.

The most accurate and precise independent measure of a star's mean 
density comes from asteroseismology by measuring the large frequency spacing 
between the pulsation modes \citep{ulrich:1986}. Thanks to \emph{Kepler's} 
precise and stable short cadence (SC) photometry \citep{gilliland:2010}, 
observers have derived fundamental properties for dozens of targets hosting 
planetary candidates \citep{huber:2013}. It is worth noting that since giants 
and sub-giant stars have greater pulsation amplitudes and timescales, 
\emph{Kepler} targets with asteroseismology include considerably more low 
surface gravity stars than a random \emph{Kepler} subset. This means that by 
using AP on this asteroseismology sample, we not only focus on the most 
well-characterized host stars, but we also have an opportunity to compare the 
ensemble population of planetary candidates associated with dwarfs versus giants.

In this work, we aim to demonstrate the value of AP in studying and 
characterizing transiting planetary candidates, where we limit our sample to
only those stars with asteroseismically determined mean stellar densities.
Since our sample contains a considerable number of evolved host stars, we will
also take this opportunity to use AP to compare the population of planetary
candidates associated with dwarfs versus giants. Our analysis represents 
the first ensemble application of AP, although we note other authors have 
conducted ensemble analyses on \emph{Kepler} candidates without using AP
(e.g. \citealt{plavchan:2014}). In \S\ref{sec:methods}, we 
outline our sample and our methodology for conducting our survey using AP. In 
\S\ref{sec:results}, we present the results of these efforts, including a
comparison between the dwarf and giant population. In \S\ref{sec:discussion}, we 
explore the possible value of AP for future missions and surveys as 
well as the implications of a high false-positive rate around giant stars.

\section{METHODS}
\label{sec:methods}

\subsection{Sample Selection}
\label{sub:sample}

From the several thousand \emph{Kepler} Objects of Interest (KOIs) known at the
time of writing, we focus on those KOIs with asteroseismically determined 
stellar densities. Whilst asteroseismology is not a requisite for conducting
AP, it is usually the most accurate and precise measurement available and is
often referred to as the ``gold standard'' \citep{bastien:2013}. 
\citet{huber:2013} recently provided a homogeneous catalog of 
asteroseismically determined stellar densities for 77 planet-candidate
host stars. This sample, including 107 planetary candidates, forms the input
catalog for our work. We subsequently refer to the independent measure of the 
stellar density used in our AP analysis as \rhoA.

In this work, we limit our sample to only those host stars with a single
transiting planet candidate. The rationale for this choice is two-fold.
First, one of the objectives of our work is to provide insights into the
false-positive rate between the dwarf and giant planet-hosting stars and
since multi-planet systems are known to have a very low false-positive rate
\citep{lissauer:2012,lissauer:2014}, the single KOIs represent the more unknown 
subset. The second reason is that we propose that the single KOIs have a higher 
a-priori probability of exhibiting large AP discrepancies than the multiple 
planet systems. This choice can be understood by considering the six known AP
effect discussed recently in \citet{AP:2014}:

\begin{itemize}
\item[{\tiny$\blacksquare$}] Photo-eccentric (PE) effect: Orbit of the 
transiting body is non-circular; causes \rhoO$>$\rhoA\ if $0<\omega<\pi$ and
\rhoO$<$\rhoA\ if $\pi<\omega<2\pi$
\item[{\tiny$\blacksquare$}] Photo-blend (PB) effect: A 
background, foreground or associated star dilutes the transit depth; causes 
\rhoO$<$\rhoA
\item[{\tiny$\blacksquare$}] Photo-timing (PT) effect: Unaccounted Transit 
Timing Variations (TTVs) affect the composite transit light curve; causes 
\rhoO$<$\rhoA
\item[{\tiny$\blacksquare$}] Photo-duration (PD) effect: Unaccounted Transit 
Duration Variations (TDVs) affect the composite transit light curve; causes 
\rhoO$<$\rhoA
\item[{\tiny$\blacksquare$}] Photo-spot (PS) effect: Unocculted starspots
behave like an anti-blend, enhancing the transit depth; causes 
\rhoO$>$\rhoA
\item[{\tiny$\blacksquare$}] Photo-mass (PM) effect: The mass of the transiting
body is significant and one cannot assume $M_{\mathrm{transiter}}\ll M_{\star}$;
causes \rhoO$>$\rhoA
\end{itemize}

Although we direct the reader to \citet{AP:2014} for exact formulae and details
of each AP effect, we point out that the PT, PD, PS and PM effects are all
generally much weaker (typically $\lesssim10^{-1}$-$10^{0}$ effect) than the PB 
and PE effects (typically $\lesssim10^{1}$-$10^{3}$). Since large AP 
discrepancies are the most interesting to study, one should expect such 
variations to be caused by either the PB or PE effects. Multi-planet 
systems certainly have a low a-priori probability of exhibiting the PB effect, 
since the false-positive rate is known to be very low \citep{lissauer:2012,
lissauer:2014}. They are also unlikely to have planets on large eccentricities 
in order to be dynamically stable, which means we expect low PE effects. For 
these reasons, we argue that single KOIs have a higher a-priori probability of 
exhibiting large and thus interesting AP discrepancies.

An added bonus of studying the single KOIs exclusively is that they are less
probable, a-priori, to exhibit TTVs and thus the PT effect. Physically speaking,
this is because they are less likely to have nearby planets near mean motion
resonance inducing large perturbations \citep{agol:2005,holman:2005}. 
\citet{mazeh:2013} recently reported that multi-transiting KOIs exhibit
significant TTVs in 120 out of 894 cases ($\simeq$13\%), whereas 
single-transiting KOIs show significant TTVs in just 23 out of 1066 cases
($\simeq$2\%). Since periodic TDVs are usually associated with periodic TTVs
\citep{kipping:2009,nesvorny:2013}, then the a-priori probability of the PD
effect is also significantly less. On this basis, we argue that any large
observed AP variations found in this sample are likely due to either i) the
photo-eccentric effect, ii) the photo-blend effect or iii) the target with
asteroseismology modes detected is not the same as the target with the 
transiting body.

In the \citet{huber:2013}, there are 43 KOIs in single transiting systems. Of
these, we exclude two objects for different reasons. KOI-42.01 shows strong
TTVs \citep{eylen:2013} and so falls in that small 2\% category of dynamically 
active single KOIs. In addition, we also exclude KOI-981.01 as our 
attempts to fit the light curve were unable to retrieve a converged, unimodal
ephemeris due to excessive correlated noise. This leaves us with 
41 KOIs for our survey, which are listed in the first column of 
Table~\ref{tab:fittedtable}. Note that we include KOIs regardless as to whether 
they have been dispositioned as a false-positive or not, since the theory of AP 
is general for any eclipsing body, not just planets \citep{AP:2014}.

\subsection{Detrending \& Fitting the Transits}
\label{sub:fitting}

We here describe the procedure used to detrend and fit the \emph{Kepler}
transit light curves. We first downloaded all available light curves spanning
quarters 1-16 for each object from the Mikulski Archive for Space Telescopes
(MAST) database\footnote{See 
http://archive.stsci.edu/kepler/data\_search/search.php}. Where available we
use the short-cadence (SC) data over the long-cadence (LC) and we always use
the Simple Aperture Photometry (SAP) time series. We exclude all data greater
than three transit durations either side of the times of transit minimum.

For each KOI, our goal is to derive posterior distributions for the fitted 
parameters in a Bayesian framework. This is achieved by coupling a Bayesian
regression routine to a transit light curve model. To model the transits, we
make the same idealized assumptions as \citet{seager:2003} e.g. spherical
planet, spherical star, circular orbits, no blended light, opaque planet, etc.
The transit light curves are generated using the \citet{mandel:2002} algorithm
described by seven free parameters with the following priors:

\begin{itemize}
\item[{\tiny$\blacksquare$}] $\log_{10}(\rho_{\star,\obs}\,[\mathrm{kg}\,\mathrm{m}^{-3}])$: 
log-base-ten of the observed stellar density. Uniform prior from 
$0<\log_{10}(\rho_{\star,\obs}\,[\mathrm{kg}\,\mathrm{m}^{-3}])<6$.
\item[{\tiny$\blacksquare$}] $(R_P/R_{\star})=p$: ratio of the planetary 
candidate's radius to the star's radius. Uniform prior $0<(R_P/R_{\star})<1$.
\item[{\tiny$\blacksquare$}] $b$: impact parameter of the transit. Uniform 
prior $0<b<2$.
\item[{\tiny$\blacksquare$}] $P$: orbital period of the planet. Uniform prior 
$(\bar{P}-1\,[\mathrm{d}])<P<(\bar{P}+1\,[\mathrm{d}])$, where $\bar{P}$ is 
the period reported by \citet{borucki:2011}.
\item[{\tiny$\blacksquare$}] $\tau$: time of transit minimum. Uniform prior 
$(\bar{\tau}-1\,[\mathrm{d}])<\tau<(\bar{\tau}+1[\,\mathrm{d}])$, where 
$\bar{\tau}$ is the period reported by \citet{borucki:2011}.
\item[{\tiny$\blacksquare$}] $q_1$: First modified quadratic limb darkening 
coefficient defined in \citet{LDfitting:2013}. Uniform prior $0<q_1<1$.
\item[{\tiny$\blacksquare$}] $q_2$: Second modified quadratic limb darkening 
coefficient defined in \citet{LDfitting:2013}. Uniform prior $0<q_2<1$.
\end{itemize}

We highlight that the stellar density is fitted uniformly in log-space, which
can also be thought of as a Jeffreys prior in $\rho_{\star,\obs}$ 
\citep{jeffreys:1946}. We choose a Jeffreys prior for this term since it can 
span several orders of magnitude and it is generally considered the most
uninformative prior choice possible. We also highlight that the limb darkening
coefficients use the uninformative priors proposed in \citet{LDfitting:2013},
which both improves the sampling efficiency and ensures complete coverage of
the physically permissible prior volume. Long-cadence data are resampled to
account for smearing using $N_{\mathrm{resam}}=30$ and the technique described
in \citet{binning:2010}.

In general, our Bayesian regression routine is a Markov Chain Monte Carlo (MCMC)
algorithm (see \citealt{gregory:2005} for the use of MCMC in uncertainty 
estimates) using the Metropolis-Hastings rule \citep{metropolis:1953,
hastings:1970}. We use a Gaussian likelihood function in all fits and the 
seven parameters are allowed to vary with jump sizes tuned to be between 
10\% and 100\% of the parameter uncertainties with the goal of $\simeq10\%$-$40$\% 
of trials accepted. For each trial, we first compute the trial model and then
determine the best fitting linear slope for each transit epoch which matches the
observations and the trial model. This is achieved using a least squares linear 
minimization routine at every trial and naturally for cases where there are many 
transits and SC data, the computational time to achieve this is significant (see 
\citealt{kundurthy:2013} for a previous example of this technique). 
Nevertheless, this approach essentially detrends the data simultaneously to
fitting the actual transit model and thus the parameter uncertainties are
more realistic.

In certain cases, visual inspection of the light curves revealed that
linear detrending was insufficient to adequately correct the 
photometry, since substantial curvature existed in the out-of-transit baseline 
data. These cases were usually, but not exclusively, long-period KOIs, 
since such transits have longer transit durations and thus the baseline can 
cover several days. In these cases, we opted to use a more sophisticated 
detrending algorithm devised by the Hunt for Exomoons with Kepler (HEK) project 
\citep{hek:2012} known as \cofiam\ (see \citealt{hek:2013} for details). This 
algorithm is well suited for longer-period planets and works on a Fourier-basis 
to guarantee that the transit profile remains undisturbed by the detrending 
procedure. The detrended light curves are then fitted using a multimodal nested 
algorithm, \multi\ \citep{feroz:2008,feroz:2009}, rather than MCMC, since 
\multi\ is both more expedient in low-dimensional space and requires no tuning 
of jump sizes.

\begin{figure}
\begin{center}
\includegraphics[width=8.4 cm]{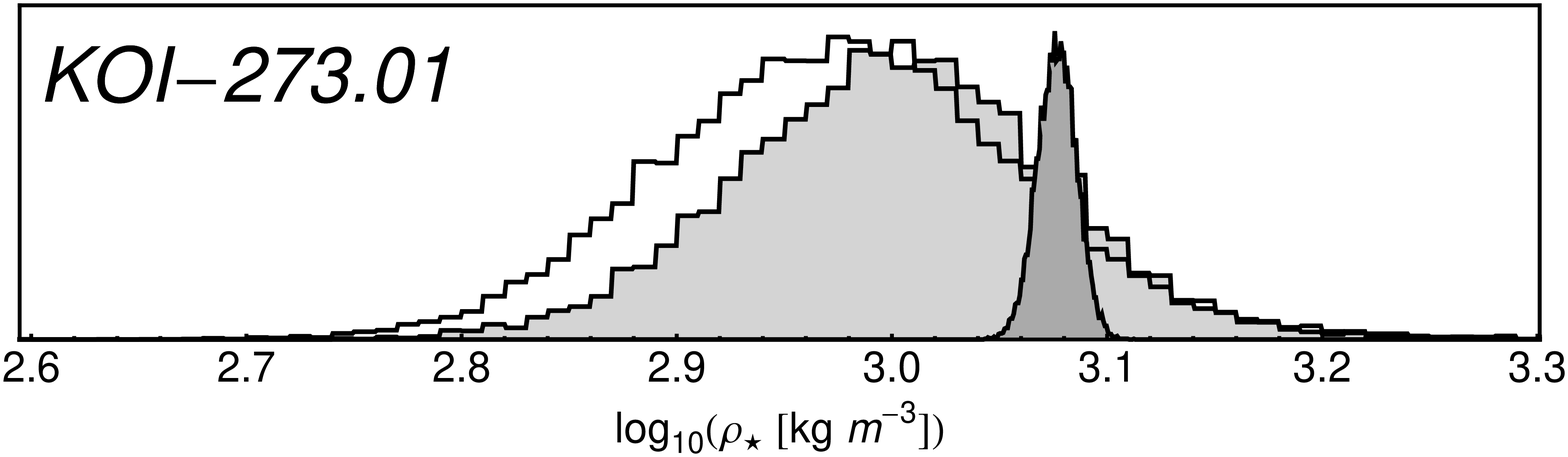}
\caption{\emph{Comparison of the posterior distribution for the light curve
derived stellar density, $\rho_{\star,\obs}$, computed using \cofiam\ plus
\multi\ (black outline with white fill) versus linear slope plus MCMC (light
gray). 
Dark gray histogram shows the asteroseismic posterior for reference.
}} 
\label{fig:KOI273}
\end{center} 
\end{figure}

It is important to note that the regression algorithms are both
well-established Bayesian Monte Carlo routines with identical input priors and 
thus the inferred posterior distributions are statistically equivalent. In cases 
where linear detrending is sufficient, one should expect consistent results 
between \cofiam\ plus \multi\ versus linear detrending plus MCMC. This was 
verified in the example of KOI-273.01, where we infer a nearly-identical 
a-posteriori distribution for $\rho_{\star,\obs}$, as shown in 
Figure~\ref{fig:KOI273}.

\section{RESULTS}
\label{sec:results}

\subsection{Conducting AP}
\label{sub:conductingAP}

\begin{figure*}
\begin{center}
\includegraphics[width=16.8 cm]{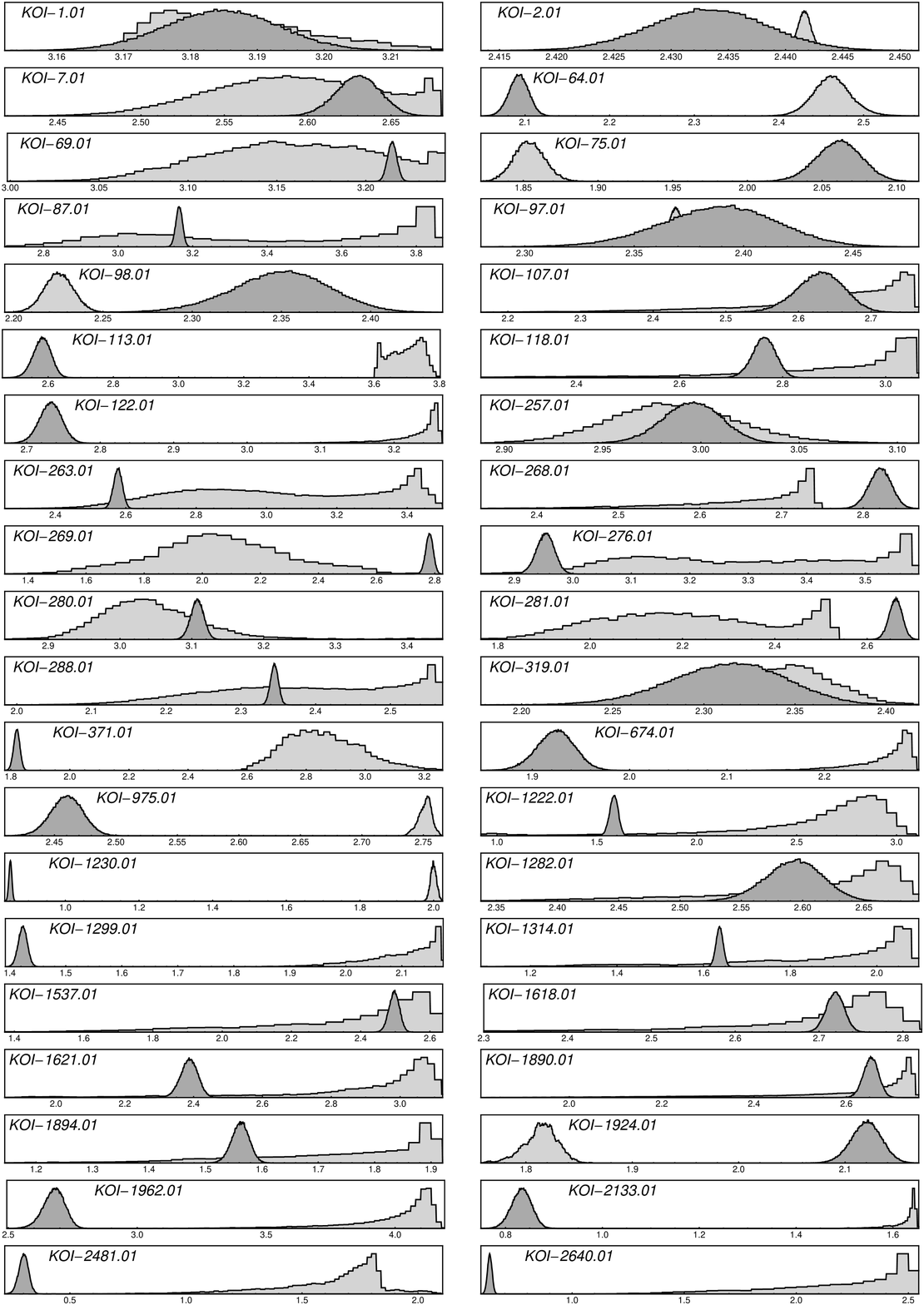}
\caption{\emph{Observed (light gray) versus asteroseismically determined
(dark gray) mean stellar density posterior distributions for the KOIs in our 
sample. The $x$-axis denotes
$\log_{10}(\rho_{\star}\,[\mathrm{kg}\,\mathrm{m}^{-3}])$. KOI-273.01 is
not included but shown in Figure~\ref{fig:KOI273}.
}} 
\label{fig:rhohistos}
\end{center} 
\end{figure*}

We provide all of the fitted transit parameters for the 41 KOIs in our sample in 
Table~\ref{tab:fittedtable}, except for $\tau$ as this is the least relevant
term for our study (available upon request). Table~\ref{tab:rhotable} provides a 
direct comparison of \rhoO\ and \rhoA, which is essentially the act of AP (this
is also visualized in Figure~\ref{fig:rhohistos}). As discussed in 
\S\ref{sub:sample}, any significant AP discrepancies in this sample are likely 
due to either the PE effect, the PB effect or that the transiting body in fact 
orbits an alternative star to the asteroseismically measured target. For 
each of these three possible explanations, we can quantify a relevant 
descriptive parameter. For the PE effect, the minimum eccentricity, 
$e_{\mathrm{min}}$, naturally falls out of the AP expressions and we use
Equation~39 of \citet{AP:2014}, the results of which are shown in Column~4 of 
Table~\ref{tab:rhotable}. Similarly, for the PB effect we use Equation~17 of 
\citet{AP:2014}, with results shown in Column~5 of Table~\ref{tab:rhotable},
provided $b<(1-p)$ which is a underlying assumption for the derivation of the PB
effect.

\input{fittedtable.tex}

If the transiting body orbits a different star in the same aperture, then we
know a) the transit is diluted and so the photo-blend effect is occurring b)
we do not know the ``true'' stellar density. However, in the case of the 
photo-blend effect, \citet{AP:2014} showed that the observed density can only be 
altered up to a minimum limit defined as (Equation~13 of \citealt{AP:2014}):

\begin{align}
\Big( \frac{\rho_{\star,\obs}}{\rho_{\star,\tru}} \Big) \geq \Bigg( \frac{ 2 p_{\obs} (1+\sqrt{1-b_{\obs}^2})}{(1+p_{\obs})^2 - b_{\obs}^2} \Bigg)^{3/2},
\end{align}

where $p_{\obs}$ and $b_{\obs}$ are the observed ratio-of-radii and impact
parameter, respectively.
In the case where $\rho_{\star,\tru}$ is unknown then, we can re-write this as

\begin{align}
\rho_{\star,\mathrm{alt,max}} = \rho_{\star,\obs} \Bigg( \frac{ 2 p_{\obs} (1+\sqrt{1-b_{\obs}^2})}{(1+p_{\obs})^2 - b_{\obs}^2} \Bigg)^{-3/2}.
\label{eqn:rhoalt}
\end{align}

Equation~\ref{eqn:rhoalt} reveals the maximum allowed density of the alternative
star. Since it is only an upper limit, only cases where 
$\rho_{\star,\mathrm{alt,max}}$ is very low allow us to exclude this scenario.
We calculate this term for every KOI in Column~6 of Table~\ref{tab:rhotable}.
In the last column, we provide a plausible list of which of these three effects,
or no effect at all (``N''), can explain the observation. Note that objects for
which $B<1$ (where $B$ is the blend factor defined in \citealt{AP:2014}) 
correspond to an anti-blend, which could be the consequence of the PS 
effect \citep{AP:2014}. Nevertheless, the PS effect is expected to cause AP 
deviations $\lesssim10$\% and thus generally does not have a significant 
impact. Additionally, although the photo-eccentric effect can explain a diverse 
range of AP observations, some cases can be rejected as being due to 
eccentricity if the periastron passage goes inside the star. We define and 
compute this using:

\begin{align}
(r_{\mathrm{peri}}/R_{\star}) &\leq (a/R_{\star}) (1-e_{\mathrm{min}}), \nonumber \\
\qquad &\leq \Bigg(\frac{\rho_{\star,\mathrm{astero}} G P^2}{3\pi}\Bigg)^{1/3} (1-e_{\mathrm{min}}).
\label{eqn:perieqn}
\end{align}

\input{rhotable.tex}

\subsection{Ensemble Results}
\label{sub:ensemble}

\begin{figure}
\begin{center}
\includegraphics[width=8.4 cm]{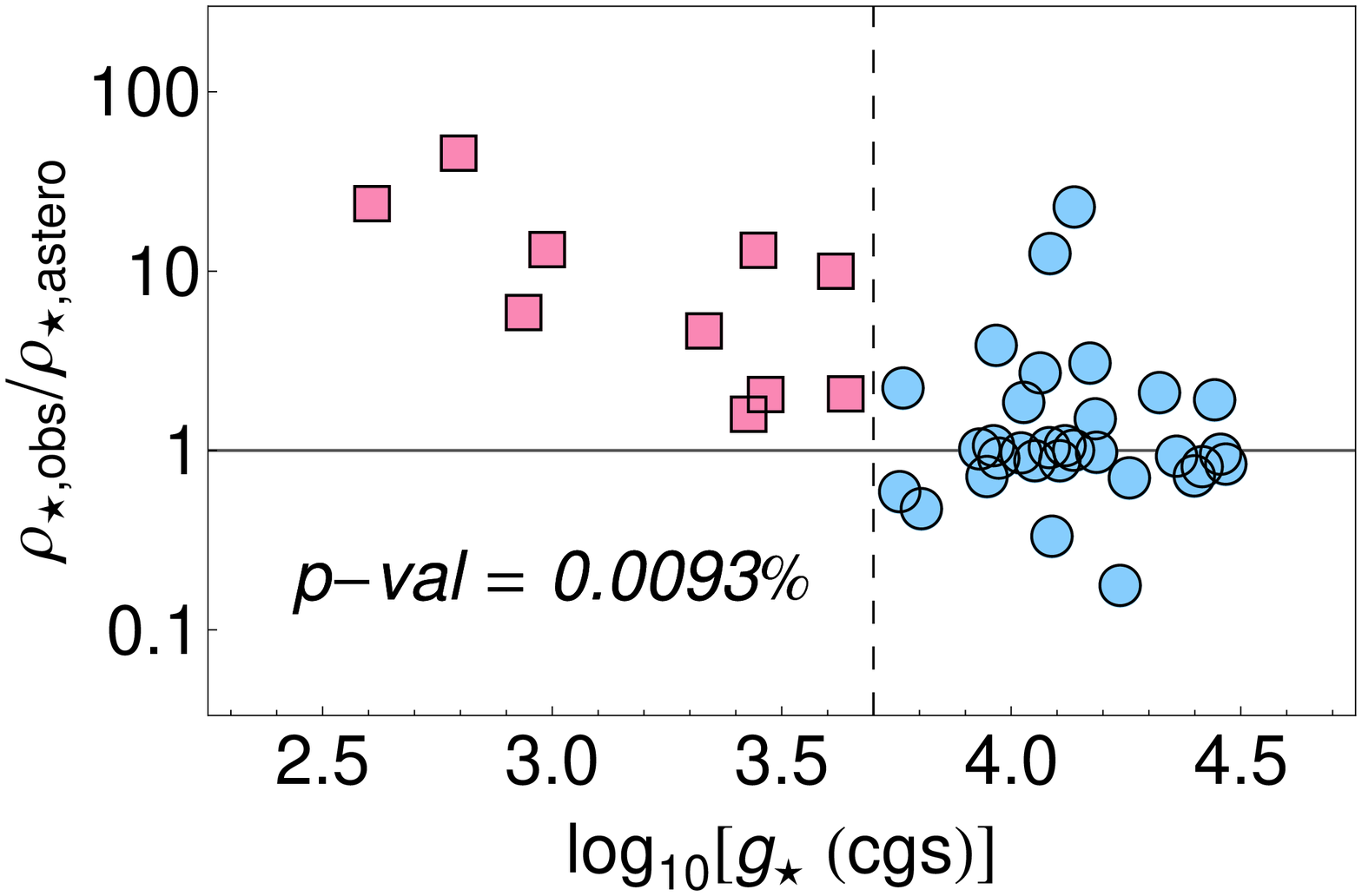}
\caption{\emph{Measured $(\rho_{\star,\obs}/\rho_{\star,\mathrm{astero}})$ from
this survey ($y$-axis) as a function of the associated star's $\log g$. Squares 
correspond to measurements with $\log g\leq3.7$ and circles for $\log g>3.7$. 
The p-value comes from an A-D test between the two populations, showing a
significant difference.
}} 
\label{fig:logg_split}
\end{center} 
\end{figure}

As evident in Table~\ref{tab:rhotable}, our AP survey of 41 single KOIs reveals 
numerous strong discrepancies between \rhoO\ and \rhoA. By plotting the 
observed discrepancies as a function of $\log g$
(values taken from \citealt{huber:2013}), one immediately identifies
an apparent split in the distribution at the boundary of $\log g = 3.7$ (see
Figure~\ref{fig:logg_split}). We choose $\log g=3.7$ as our split since this is 
a reasonable proxy for the boundary between dwarfs and giants/sub-giants, plus 
all of the observed densities below this limit have an overestimated value.
Performing an \citet{AD:1952} test between these two populations 
reveals a 0.009\% chance they are drawn from the same underlying distribution.

The evidence for a significant divide is supported by further analysis too.
We decided to investigate how well our measured 
$(\rho_{\star,\obs}/\rho_{\star,\mathrm{astero}})$ distribution matches that
which would be expected if the photo-eccentric effect alone was responsible
for the observations. The motivation for this is that the photo-eccentric
effect is capable of explaining the greatest range of measurements 
\citep{AP:2014}. We therefore generated a synthetic population of 
$(\rho_{\star,\obs}/\rho_{\star,\mathrm{astero}})$ measurements using:

\begin{align}
\Big( \frac{\rho_{\star,\obs}}{\rho_{\star,\tru}} \Big) &= \frac{(1+e\sin\omega)^3}{(1-e^2)^{3/2}}.
\label{eqn:Psi}
\end{align}

\begin{figure*}
\begin{center}
\includegraphics[width=18.0 cm]{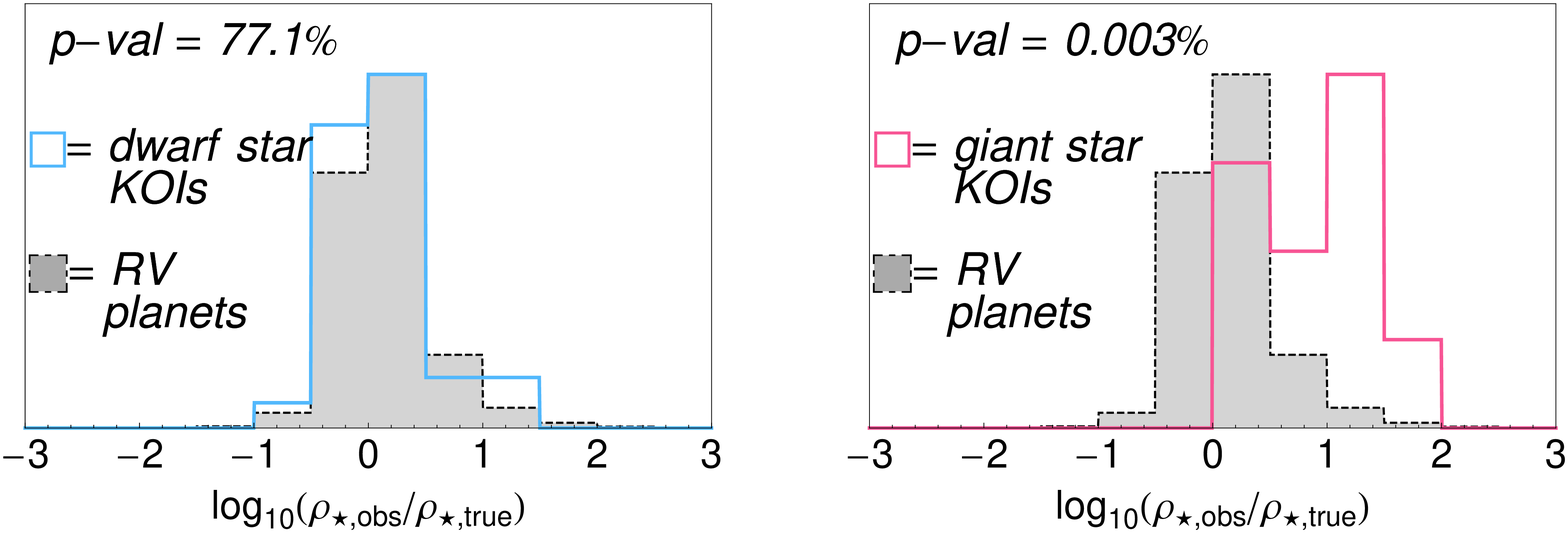}
\caption{\emph{Histograms of $(\rho_{\star,\obs}/\rho_{\star,\mathrm{astero}})$
for KOIs studied in this work. On the left we show the results for those KOIs
orbiting dwarf stars ($\log g>3.7$) and on the right those for giant stars
($\log g\leq3.7$), demonstrating the clear difference between the two
subsets. The gray histogram shows that which would be expected if only the
photo-eccentric effect was occurring and the eccentricity distribution matched
that observed from the radial velocity planets \citep{beta:2013} deliberately
binned to the same scale.}} 
\label{fig:APcomp}
\end{center} 
\end{figure*}

Using the code \eccsamples\ \citep{ECCSAMPLES:2014}, we draw random $e$ and 
$\omega$ samples from the joint probability distribution 
$\mathrm{P}(e,\omega|\mathrm{object\,\,known\,\,to\,\,transit})$ 
to generate our synthetic photo-eccentric population. \eccsamples\ assumes a 
Beta distribution for the underlying probability distribution
of $e$, $\mathrm{P}(e)$ with shape parameters $a$ and $b$. In our simulation,
we adopt $a=0.867$ and $b=3.03$, which has been shown to provide an excellent
match to the observed eccentricity distribution from radial velocity surveys
\citep{beta:2013}. The final synthetic 
$(\rho_{\star,\obs}/\rho_{\star,\mathrm{astero}})$ distribution is shown in
Figure~\ref{fig:APcomp} as a gray histogram. Overlaying the observed
distributions for the dwarfs and giants immediately demonstrates how different
the two samples are. The dwarf sample is fully compatible with the 
photo-eccentric effect, with a A-D test giving a $p$-value of $77.1\%$. In
contrast, the giants are grossly incompatible with the photo-eccentric
effect, with a A-D $p$-value of $0.003$\%, or 4.2\,$\sigma$.

In considering this puzzling observation, we devised four possible hypotheses
which could reconcile this split:

\begin{enumerate}
\item The larger pulsations of the giants induce significant time-correlated
noise in the folded light curve, which subsequently skews the 
$\rho_{\star,\obs}$ determination.
\item The asteroseismically determined densities systematically 
underestimate the stellar density for giant stars.
\item Companions to giant stars are highly eccentric and have a dramatically
different eccentricity distribution than dwarf stars.
\item A large fraction of the KOIs associated with giant stars in fact orbit
a different star within the aperture - cases we define as false-positives.
\end{enumerate}

Hypothesis 1 can be tested by first quantifying the degree of time-correlated
noise in the data. The timescale of this spurious noise must have dominant
power at $\nu_{\mathrm{max}}$, the frequency of maximum asteroseismology
power. For many of the targets, \citet{huber:2013} directly provide 
$\nu_{\mathrm{max}}$ and where unavailable we use Equation~10 of 
\citet{kjeldsen:1995}. We then tried cross-correlating the 
$(\rho_{\star,\obs}/\rho_{\star,\mathrm{astero}})$ measurements to the
transit duration normalized by this timescale. Performing an A-D test about the
median of this new variable, as we did with $\log g$ before, finds no 
significant split with a $p$-value of 3\%. We also repeated this exercise
using transit depth normalized by the amplitude of the maximum pulsation
(computed using Equation~8 of \citealt{kjeldsen:1995}) and this yields a
$p$-value of 14\%. If time-correlated noise was genuinely responsible, one 
should expect these $p$-values should be lower than that found when using 
$\log g$ as the variable. Finally, we note that the median period of the giant 
sample is 25\,d and thus the number of transits stacked together is typically 
large. This folding effect reduces the effect of time-correlated noise as 
$N_{\mathrm{transits}}^{-1/2}$ \citep{pont:2006}, further detracting from 
hypothesis 1. We therefore conclude hypothesis 1 is an improbable explanation 
for the observed distribution.

Hypothesis 2 seems improbable on the basis that giant stars yield
the largest pulsations amplitudes and timescales. Further, the density of
the star is typically the most precisely determined parameter from 
asteroseismology, directly related to the frequency splitting, $\Delta\nu$ via
\citep{ulrich:1986}:

\begin{align}
\Delta \nu &= \frac{ (M_{\star}/M_{\odot})^{1/2} }{ (R_{\star}/R_{\odot})^{3/2} } \Delta\nu_{\odot}.
\end{align}

The possibility that companions to giant stars are highly eccentric, hypothesis
3, has no direct physical motivation. The median \emph{minimum} eccentricity 
required to explain this observation is $0.60$. This appears inconsistent with 
the planets detected $\log g\leq3.7$ stars from radial velocities, for which the 
median eccentricity is much lower at $0.129$ (see www.exoplanets.org 
\citealt{wright:2011}). Secondly, even though the photo-eccentric effect is
expected to yield a small overestimate bias \citep{AP:2014}, the fact 
that none of the objects have an underestimation effect is improbable.
Finally, four of the ten giants (KOI-1222.01, KOI-2133.01, 
KOI-2481.01 \& KOI-2640.01) cannot possibly be explained by the
photo-eccentric effect, since this requires a periastron passage inside the 
star, which we deem unphysical. We therefore find the super-eccentric planets
hypothesis strongly disfavored.

By deduction, this leaves us with hypothesis 4 as the only viable explanation.
As discussed in \S\ref{sec:discussion}, this appears consistent with independent
arguments regarding the false-positive rate for this sample. Given the small
number statistics involved with a sample of just 10 giant star KOIs, the 
precision to which the associated false-positive rate (FPR) can be measured is
naturally low. However, from Figure~\ref{fig:APcomp}, we estimate that
three of the ten giant star KOIs have 
$(\rho_{\star,\obs}/\rho_{\star,\mathrm{astero}})$ compatible with the
synthetic photo-eccentric effect population. The other seven have sufficiently 
high $(\rho_{\star,\obs}/\rho_{\star,\mathrm{astero}})$ measurements that they
appear incompatible with the synthetic photo-ecentric population. On this
basis, we estimate FPR\,$\simeq(70\pm30)$\%. We further note that the FPR can
easily be seen to be at least FPR$\gtrsim(40\pm20)$\%, on the basis that four of
the ten KOIs are classified unambiguously as false-positives in 
Table~\ref{tab:rhotable}, since they would have to be so eccentric they would 
pass inside or contact the star.

\subsection{Kepler-91b: A False-Positive?}
\label{sub:kepler91}

Recently, \citet{lillo-box:2013} claimed to confirm the planetary nature of
KOI-2133.01, or Kepler-91b, which is an object in our sample. If this KOI
was genuinely a planet, it would seem to be a counter-example to our conclusion
of a high false-positive rate for giant host stars. Further, in 
Table~\ref{tab:rhotable} we identify KOI-2133.01 as an unambiguous FP using AP. 
For these reasons, it is important that we investigate this apparent 
discrepancy.

The key reason why we identified this object as a false-positive is because
$\rho_{\star,\obs}$ is so much larger than $\rho_{\star,\mathrm{astero}}$ that
the orbit would have to be highly eccentric, such that 
$(r_{\mathrm{peri}}/R_{\star})=1.10_{-0.05}^{+0.06}$ i.e. the planet is
essentially in-contact with the star. Specifically, we have
$\rho_{\star,\mathrm{astero}}=6.81\pm0.032$\,kg\,m$^{-3}$ but 
$\rho_{\star,\obs}=43.47_{-3.35}^{+0.67}$\,kg\,m$^{-3}$, which may be 
equivalently expressed in terms of the semi-major axis using Kepler's Third Law 
as $(a/R_{\star})_{\obs} = 4.476_{-0.118}^{+0.023}$. Critically, 
\citet{lillo-box:2013} find a much lower observed stellar density, which is more 
compatible with $\rho_{\star,\mathrm{astero}}$ and thus does not require a 
highly eccentric planet. They report 
$\rho_{\star,\obs}=7.1_{-1.9}^{+0.7}$\,kg\,m$^{-3}$, which is equivalent
to $(a/R_{\star})_{\obs} = 2.32_{-0.22}^{+0.07}$.

With two dramatically different light curve determinations of 
$\rho_{\star,\obs}$, or equivalently $(a/R_{\star})_{\obs}$, it remains unclear 
which solution is correct. Fortunately, two additional independent studies have 
also computed light curve solutions for this object, namely \citet{burke:2013} 
and \citet{esteves:2013}. In the case of \citet{burke:2013}, Table~1 reports 
$(a/R_{\star}) = 4.346$ (no associated uncertainty reported), which is within 
1.1\,$\sigma$ of our solution but $>25$\,$\sigma$ discrepant to that of 
\citet{lillo-box:2013}. In the case of \citet{esteves:2013}, the authors report 
$(a/R_{\star}) = 4.51_{-0.26}^{+0.12}$, which is in excellent agreement with our 
solution ($<1$\,$\sigma$) and inconsistent with that of \citet{lillo-box:2013} 
($>8$\,$\sigma$). We note that \citet{lillo-box:2013} cite 
\citet{tenenbaum:2013} as finding $(a/R_{\star}) = (2.64\pm0.23)$, however this 
value is not actually listed anywhere is \citet{tenenbaum:2013} and the authors 
have stated this is not a result from their paper (P. Tenenbaum; 2014 private 
communication).

Additionally, \citet{esteves:2013} also identified KOI-2133.01 as a 
false-positive using a completely different technique than us. They reported 
strong phase variations indicative of reflected light and ellipsoidal 
variations. However, if $(a/R_{\star}) \simeq 4.5$, the amplitude of the 
variations is so great that KOI-2133.01 must be self-luminous and thus a 
false-positive. \citet{lillo-box:2013} remark on this but since their 
$(a/R_{\star})$ value is much lower, the phase variations can be explained by
reflected light without KOI-2133.01 being self-luminous.

We point out that KOI-2133.01 has a short orbital period ($6.25$\,days) giving 
us 221 transits which we fitted in this work. In general, one does not expect
time-correlated noise to phase up coherently when the transits are folded upon
a linear ephemeris \citep{pont:2006}. For this reason, the large number of
transits for KOI-2133.01 should lead to red noise being heavily attenuated via 
$1/\sqrt{N}$, where $N$ is the number of transits. For this reason and the 
reasons discussed in \S\ref{sub:ensemble}, it would be surprising if red noise 
could be responsible for a erroneous $(a/R_{\star})$.

It should therefore be clear that the planetary-nature of KOI-2133.01 hangs
primarily as to whether $(a/R_{\star}) \simeq 4.5$, in which
case it is a false-positive, or $(a/R_{\star}) \simeq 2.3$, in which case it can
be a planet. With three independent measurements by ourselves, 
\citet{burke:2013} and \citet{esteves:2013} in agreement versus one study
finding the planet scenario \citep{lillo-box:2013}, the current consensus would
favor the false-positive scenario. However, we would encourage multiple
independent groups to study this light curve in order to resolve this important
question.

\section{DISCUSSION \& CONCLUSIONS}
\label{sec:discussion}

\subsection{Comparing Our FPR to the Literature}
\label{sub:litcomp}

In this work, using the novel technique of AP in isolation, we demonstrate that
the false-positive rate of \emph{Kepler} planetary candidates associated with
stars of $\log g\leq3.7$ (i.e. the giants and sub-giants) is 
FPR$\simeq70\%\pm30$\% (see \S\ref{sub:ensemble}). Due to the small-number
statistics of our giant-star sample, we prefer the interpretation of a merely
``high'' false-positive rate, rather than an explicit numerical value.
In contrast, we find no compelling evidence for a non-zero FPR of the 
$\log g>3.7$ sample (i.e. the dwarfs). This latter result is consistent with the 
low FPR for \emph{Kepler} dwarfs reported by \citet{morton:2011} and 
\citet{fressin:2013} of $\lesssim10$\%. However, we are aware of no previous 
studies characterizing the FPR for \emph{Kepler's} giant star population.

\begin{figure}
\begin{center}
\includegraphics[width=8.4 cm]{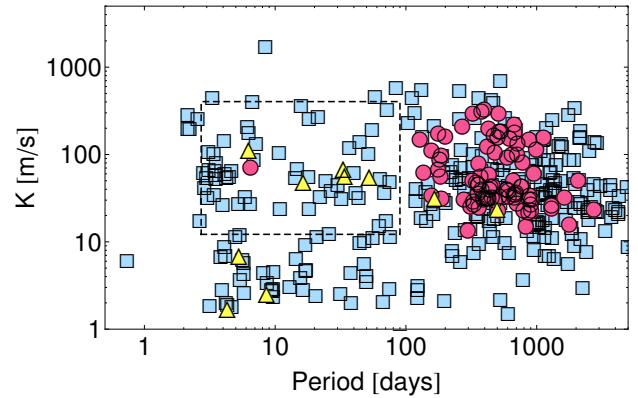}
\caption{\emph{
Radial velocity semi-amplitudes as a function of the planetary
period. Squares denote known planets around host stars with $\log g>3.7$ and
circles denote those for $\log g\leq3.7$. Triangles represent the estimated
position of the giant star KOIs in our sample (not measured radial velocities).
The occurrence rate of giant star planets in the dashed box area is very low, 
with only one known member, HD~102956b \citep{johnson:2010}. The location of
five triangles in this region is therefore compatible with a high FPR for giant 
stars. Data come from www.exoplanets.org \citep{wright:2011}.
}} 
\label{fig:RVcomp}
\end{center} 
\end{figure}

Although there are no explicit studies regarding \emph{Kepler's} giant star FPR, 
numerous other works indicate our result is not a surprise. For example, the 
population of exoplanets discovered using the radial velocity (RV) technique 
provides some useful insights. In Figure~\ref{fig:RVcomp}, it is apparent that 
there is a paucity of planets detected with periods below 100\,days for host 
stars with $\log g<3.7$ ($1/87$), where the data come from www.exoplanets.org 
\citep{wright:2011}. Eight of the ten giant planetary candidates studied 
in our sample have periods below 100\,days though, and therefore seem to occupy a 
parameter space where radial velocities predict a low occurrence rate.
To investigate this further, we estimated the approximate radial velocity 
amplitudes of the ten KOIs. Approximate masses were estimated from the radii 
using the empirical relation of \citet{weiss:2014} for planets below 4\,Earth 
radii and a simple power-law interpolation through the www.exoplanets.org 
catalog \citep{wright:2011} for larger worlds, capping masses off at 1.3\,$M_J$.
The estimated RV amplitudes are shown as triangles in Figure~\ref{fig:RVcomp}. 
This reveals that indeed five of the KOIs occupy a region where genuine planets 
are very rare. This lends credence to the hypothesis that the FPR for our sample 
is high.

Another useful insight comes from the number of multiple transiting planet
systems detected between the dwarfs and giants, since the FPR of multiple 
planet systems is known to be very low \citep{lissauer:2012,lissauer:2014} 
at FPR$\lesssim$1\%. Using the catalog available at
http://exoplanetarchive.ipac.caltech.edu, we count that 1.8\% of host stars 
with $\log g\leq3.7$ and with KOIs not dispositioned as a false-positive
reside in multiple transiting KOI systems. In contrast, doing the same for 
the $\log g>3.7$ sample yields 16.9\% of the objects. In other words, a 
\emph{Kepler} star identified to have transiting planetary candidates is 
nearly 10 times more likely to have multiple candidates if it is a dwarf 
rather than a giant. This test is not definitive since the number of 
multi-planet systems orbiting giant stars may genuinely be much lower,
but equally it can be explained by an order of magnitude higher false-positive 
rate for the giants.

We argue that our sample of single KOIs associated with giant stars is largely
unbiased, since the only selection criterion is the presence of detectable 
oscillation modes. This criterion implies that the target a) exhibits large 
oscillation modes b) is bright enough for these modes to be detected. Amongst 
the population of giant stars, point a) introduces no significant bias, since 
the amplitude of maximum oscillation power is enhanced for all low $\log g$ 
targets via $\sim T_{\mathrm{eff}}^2/\log g$ \citep{kjeldsen:1995}. The second 
point implies we are biased towards brighter targets and so our population may 
be closer than giant stars without asteroseismic detections. If anything, 
targets further away may have a higher false-positive rate due to an increased 
probability of chance alignment. We also note that oscillations may be damped 
for short-period binaries \citep{gaulme:2014}, however such systems are
easily identified via large ellipsoidal variations and beaming effects and
unlikely to be a significant source of bias in our sample either.

Our AP survey concludes that many of the giants are false-positives, which 
specifically is defined as meaning that the transiting body is actually
eclipsing a different star. This implies that we should expect many of these
KOIs to have potentially detectable companions using adaptive optics (AO)
imaging. To investigate this, we use the database of AO images acquired for
715 KOIs by \citet{law:2013}. Of these 715, just 19 targets have 
$\log g\leq3.7$ which include the 10 giant star KOIs in our survey ($\log g$
estimates taken from \citealt{huber:2014}). \citet{law:2013} report no 
detections of companions for any of these 19 KOIs, with typical limits of 
$\Delta m\simeq6$ from $\approx$0\farcs15 to 2\farcs5. However, AO
imaging is less constraining for the giants since they are intrinsically
brighter and thus in a magnitude-limited survey like \emph{Kepler} will
have to be at greater distance from the observer. Using the stellar parameters 
of \citet{huber:2014}, we estimate that the dwarf KOIs ($\log g>3.7$) have a 
median distance of $d=820_{-400}^{+420}$\,pc, whereas the giants 
($\log g\leq3.7$) have $d=1500_{-700}^{+2200}$\,pc. We therefore do not consider 
the lack of AO detections to be incompatible with our result.

\subsection{Future Possibilities of AP}
\label{sub:future}

This work demonstrates the unique power of the relatively new technique of
asterodensity profiling. Whilst AP is usually associated with the goal of
constraining the orbital eccentricities of exoplanets 
\citep{MAP:2012,dawson:2012}, we here verify that the method is also a powerful 
tool in vetting planetary candidates using photometry alone \citep{tingley:2011,
AP:2014}. In this work, we considered just 41 KOIs, but future studies 
with hundreds or thousands of objects would be able to realistically measure the 
eccentricity distribution using AP alone. Ensemble studies require targets with 
homogeneously and accurately derived stellar densities and we encourage work in 
this area to provide AP a larger sample of targets for future applications.

Future space-based transit survey missions, such as 
TESS\footnote{http://tess.gsfc.nasa.gov/index.html} \citep{ricker:2010}
and PLATO \citep{rauer:2013}, will also surely benefit from using AP in both 
planet validation and characterization and our work highlights the value of 
accurate stellar parameters for such surveys.

\acknowledgements
\section*{Acknowledgements}

We thank the anonymous reviewer for their helpful comments.
We offer our thanks and praise to the extraordinary scientists, engineers
and individuals who have made the \emph{Kepler Mission} possible.
D. Sliski wishes to thank the Michele and David Mittleman Foundation for their 
computational support in helping to reduce the mass amount of data. 
This research has made use of the Exoplanet Orbit Database and the Exoplanet 
Data Explorer at exoplanets.org.


\end{document}

%% file: shortcuts.tex
\newcommand{\multi}{{\sc MultiNest}}
\newcommand{\cofiam}{{\tt CoFiAM}}
\newcommand{\obs}{\mathrm{obs}}
\newcommand{\tru}{\mathrm{true}}
\newcommand{\rhoO}{$\rho_{\star,\obs}$}

\newcommand{\rhoA}{$\rho_{\star,\mathrm{astero}}$}
\newcommand{\eccsamples}{{\tt ECCSAMPLES}}

%% file: fittedtable.tex
\begin{table*}
\caption{\emph{Fitted Transit Light Curve Parameters for 41 Single KOIs
with Asteroseismology. Rows above the horizontal line have $\log g>3.7$ and
those below have $\log g \leq 3.7$.}} 
\centering 
\begin{tabular}{l l l l l l l} 
\hline\hline 
KOI & $R_P/R_{\star}$ & $\log_{10}(\rho_{\star}\,[\mathrm{kg}\,\mathrm{m}^{-3}])$ & $b$ & $P$\,[days] & $q_1$ & $q_2$ \\ [0.5ex] 
\hline
1.01 & $0.1258_{-0.0015}^{+0.0020}$ & $3.1853_{-0.0096}^{+0.0134}$ & $0.8442_{-0.0027}^{+0.0025}$ & $2.4706123_{-0.0000024}^{+0.0000023}$ & $0.344_{-0.034}^{+0.037}$ & $0.31_{-0.22}^{+0.29}$ \\
2.01 & $0.077738_{-0.000012}^{+0.000012}$ & $2.44165_{-0.00058}^{+0.00058}$ & $0.50033_{-0.00078}^{+0.00079}$ & $2.204735409_{-0.000000015}^{+0.000000015}$ & $0.2712_{-0.0014}^{+0.0014}$ & $0.3366_{-0.0028}^{+0.0028}$ \\
7.01 & $0.02505_{-0.00024}^{+0.00023}$ & $2.593_{-0.051}^{+0.055}$ & $0.351_{-0.142}^{+0.087}$ & $3.2136701_{-0.00000065}^{+0.00000065}$ & $0.385_{-0.032}^{+0.034}$ & $0.333_{-0.043}^{+0.046}$ \\
64.01 & $0.03933_{-0.00057}^{+0.00096}$ & $2.463_{-0.021}^{+0.022}$ & $0.9380_{-0.0033}^{+0.0033}$ & $1.95108246_{-0.00000026}^{+0.00000026}$ & $0.474_{-0.075}^{+0.068}$ & $0.15_{-0.11}^{+0.21}$ \\
69.01 & $0.01508_{-0.00013}^{+0.00014}$ & $3.160_{-0.051}^{+0.051}$ & $0.346_{-0.129}^{+0.086}$ & $4.72673879_{-0.00000053}^{+0.00000053}$ & $0.358_{-0.023}^{+0.025}$ & $0.386_{-0.039}^{+0.041}$ \\
75.01 & $0.03979_{-0.00012}^{+0.00012}$ & $1.8538_{-0.0099}^{+0.0105}$ & $0.6808_{-0.0068}^{+0.0063}$ & $105.881608_{-0.000034}^{+0.000034}$ & $0.311_{-0.010}^{+0.011}$ & $0.559_{-0.040}^{+0.041}$ \\
87.01 & $0.0215_{-0.0012}^{+0.0017}$ & $3.47_{-0.46}^{+0.35}$ & $0.67_{-0.48}^{+0.19}$ & $289.86442_{-0.00087}^{+0.00088}$ & $0.41_{-0.16}^{+0.32}$ & $0.16_{-0.12}^{+0.35}$ \\
97.01 & $0.082950_{-0.000083}^{+0.000084}$ & $2.3688_{-0.0033}^{+0.0032}$ & $0.5574_{-0.0036}^{+0.0037}$ & $4.88548901_{-0.00000018}^{+0.00000018}$ & $0.3062_{-0.0085}^{+0.0088}$ & $0.354_{-0.015}^{+0.016}$ \\
98.01 & $0.045650_{-0.000077}^{+0.000076}$ & $2.2252_{-0.0080}^{+0.0082}$ & $0.5893_{-0.0075}^{+0.0072}$ & $6.79012304_{-0.00000060}^{+0.00000061}$ & $0.2760_{-0.0093}^{+0.0096}$ & $0.350_{-0.025}^{+0.026}$ \\
107.01 & $0.01993_{-0.00022}^{+0.00052}$ & $2.679_{-0.145}^{+0.064}$ & $0.33_{-0.20}^{+0.21}$ & $7.2569658_{-0.0000043}^{+0.0000043}$ & $0.392_{-0.075}^{+0.090}$ & $0.35_{-0.10}^{+0.11}$ \\
113.01 & $0.75_{-0.28}^{+0.17}$ & $3.703_{-0.067}^{+0.048}$ & $1.56_{-0.30}^{+0.18}$ & $386.5980986_{-0.0000017}^{+0.0000007}$ & $0.609_{-0.083}^{+0.089}$ & $0.15_{-0.11}^{+0.22}$ \\
118.01 & $0.01549_{-0.00028}^{+0.00056}$ & $2.964_{-0.237}^{+0.077}$ & $0.34_{-0.26}^{+0.28}$ & $24.993233_{-0.000042}^{+0.000040}$ & $0.29_{-0.12}^{+0.15}$ & $0.44_{-0.25}^{+0.31}$ \\
122.01 & $0.02061_{-0.00010}^{+0.00014}$ & $3.246_{-0.046}^{+0.012}$ & $0.13_{-0.12}^{+0.16}$ & $11.5230707_{-0.0000040}^{+0.0000041}$ & $0.369_{-0.060}^{+0.067}$ & $ 0.314_{-0.079}^{+0.087}$ \\
257.01 & $0.02371_{-0.00050}^{+0.00062}$ & $2.985_{-0.031}^{+0.033}$ & $0.8621_{-0.0081}^{+0.0073}$ & $6.8834063_{-0.0000012}^{+0.0000012}$ & $0.298_{-0.023}^{+0.022}$ & $0.47_{-0.28}^{+0.28}$ \\
263.01 & $0.01466_{-0.00096}^{+0.00081}$ & $3.03_{-0.29}^{+0.36}$ & $0.70_{-0.042}^{+0.12}$ & $20.719416_{-0.000029}^{+0.000028}$ & $0.38_{-0.13}^{+0.30}$ & $0.29_{-0.22}^{+0.41}$ \\
268.01 & $0.02063_{-0.00015}^{+0.00028}$ & $2.691_{-0.118}^{+0.041}$ & $0.26_{-0.20}^{+0.21}$ & $110.37849_{-0.00013}^{+0.00013}$ & $0.339_{-0.061}^{+0.072}$ & $0.146_{-0.080}^{+0.082}$ \\
269.01 & $0.01056_{-0.00033}^{+0.00036}$ & $2.05_{-0.22}^{+0.23}$ & $0.825_{-0.089}^{+0.053}$ & $18.011628_{-0.000029}^{+0.000030}$ & $0.375_{-0.059}^{+0.067}$ & $0.045_{-0.038}^{+0.070}$ \\
273.01 & $0.0214_{-0.0011}^{+0.0017}$ & $2.959_{-0.080}^{+0.083}$ & $0.9325_{-0.0100}^{+0.0086}$ & $10.5737625_{-0.0000071}^{+0.0000072}$ & $0.608_{-0.092}^{+0.083}$ & $0.49_{-0.35}^{+0.34}$ \\
276.01 & $0.01978_{-0.00074}^{+0.00083}$ & $3.30_{-0.22}^{+0.23}$ & $0.59_{-0.36}^{+0.14}$ & $41.746004_{-0.000032}^{+0.000032}$ & $0.396_{-0.087}^{+0.143}$ & $0.26_{-0.15}^{+0.22}$ \\
280.01 & $0.01952_{-0.00031}^{+0.00045}$ & $3.036_{-0.060}^{+0.074}$ & $0.866_{-0.018}^{+0.013}$ & $11.8728958_{-0.0000044}^{+0.0000044}$ & $0.363_{-0.043}^{+0.044}$ & $0.18_{-0.14}^{+0.24}$ \\
281.01 & $0.01695_{-0.00069}^{+0.00058}$ & $2.21_{-0.19}^{+0.25}$ & $0.62_{-0.32}^{+0.11}$ & $19.556609_{-0.000023}^{+0.000023}$ & $0.392_{-0.070}^{+0.091}$ & $0.32_{-0.15}^{+0.17}$ \\
288.01 & $0.01377_{-0.00031}^{+0.00035}$ & $2.40_{-0.14}^{+0.14}$ & $0.48_{-0.27}^{+0.14}$ & $10.2753113_{-0.0000055}^{+0.0000055}$ & $0.334_{-0.047}^{+0.055}$ & $0.40_{-0.10}^{+0.12}$ \\
319.01 & $0.0485_{-0.0019}^{+0.0019}$ & $2.344_{-0.027}^{+0.025}$ & $0.9050_{-0.0042}^{+0.0041}$ & $46.151110_{-0.000026}^{+0.000026}$ & $0.430_{-0.036}^{+0.039}$ & $0.58_{-0.39}^{+0.30}$ \\
975.01 & $0.007573_{-0.000019}^{+0.000019}$ & $2.7520_{-0.0061}^{+0.0045}$ & $0.178_{-0.018}^{+0.023}$ & $2.78581687_{-0.00000063}^{+0.00000057}$ & $0.280_{-0.029}^{+0.031}$ & $0.425_{-0.053}^{+0.057}$ \\
1282.01 & $0.04729_{-0.00072}^{+0.00105}$ & $2.636_{-0.110}^{+0.037}$ & $0.23_{-0.18}^{+0.22}$ & $30.863933_{-0.000080}^{+0.000078}$ & $0.36_{-0.14}^{+0.20} $ & $0.48_{-0.20}^{+0.28}$ \\
1537.01 & $0.00720_{-0.00032}^{+0.00071}$ & $2.45957_{-0.383943}^{+0.120555}$ & $0.43_{-0.29}^{+0.32}$ & $10.191592_{-0.000049}^{+0.000045}$ & $0.47_{-0.20}^{+0.25}$ & $0.66_{-0.31}^{+0.25}$ \\
1618.01 & $0.00514_{-0.00013}^{+0.00014}$ & $2.734_{-0.147}^{+0.046}$ & $0.23_{-0.20}^{+0.28}$ & $2.3643709_{-0.0000062}^{+0.0000064}$ & $0.25_{-0.16}^{+0.25}$ & $0.25_{-0.20}^{+0.43}$ \\
1621.01 & $0.01199_{-0.00031}^{+0.00048}$ & $2.998_{-0.326}^{+0.080}$ & $0.36_{-0.27}^{+0.34}$ & $20.310507_{-0.000054}^{+0.000053}$ & $0.27_{-0.14}^{+0.29}$ & $0.18_{-0.16}^{+0.36}$ \\
1890.01 & $0.00954_{-0.00014}^{+0.00042}$ & $2.679_{-0.264}^{+0.056}$ & $0.30_{-0.22}^{+0.33}$ & $4.3364290_{-0.0000038}^{+0.0000034}$ & $0.38_{-0.11}^{+0.15}$ & $0.37_{-0.20}^{+0.23}$ \\
1924.01 & $0.004264_{-0.000056}^{+0.000065}$ & $1.815_{-0.017}^{+0.014}$ & $0.341_{-0.032}^{+0.035}$ & $2.1191569_{-0.0000021}^{+0.0000023}$ & $0.986_{-0.013}^{+0.010}$ & $0.99955_{-0.00160}^{+0.00045}$ \\
1962.01 & $0.03700_{-0.00092}^{+0.00137}$ & $4.064_{-0.221}^{+0.066}$ & $0.31_{-0.23}^{+0.29}$ & $32.858685_{-0.000055}^{+0.000051}$ & $0.36_{-0.17}^{+0.31}$ & $0.39_{-0.26}^{+0.34}$ \\
\hline
371.01 & $0.40_{-0.29}^{+0.45}$ & $2.85_{-0.12}^{+0.14}$ & $1.36_{-0.30}^{+0.45}$ & $498.3915_{-0.0012}^{+0.0012}$ & $0.33_{-0.25}^{+0.30}$ & $0.66_{-0.37}^{+0.25}$ \\
674.01 & $0.03772_{-0.00020}^{+0.00029}$ & $2.274_{-0.038}^{+0.013}$ & $0.15_{-0.11}^{+0.13}$ & $16.338893_{-0.000013}^{+0.000013}$ & $0.356_{-0.055}^{+0.062}$ & $0.511_{-0.079}^{+0.092}$ \\
1222.01 & $0.00513_{-0.00035}^{+0.00049}$ & $2.74_{-0.41}^{+0.18}$ & $0.41_{-0.28}^{+0.35}$ & $4.285768_{-0.000061}^{+0.000046}$ & $0.54_{-0.34}^{+0.32}$ & $0.53_{-0.35}^{+0.33}$ \\
1230.01 & $0.07859_{-0.00020}^{+0.00019}$ & $1.9996_{-0.0082}^{+0.0086}$ & $0.317_{-0.021}^{+0.019}$ & $165.739391_{-0.000084}^{+0.000080}$ & $0.390_{-0.017}^{+0.019}$ & $0.441_{-0.021}^{+0.021}$ \\
1299.01 & $0.02855_{-0.00028}^{+0.00057}$ & $2.118_{-0.084}^{+0.044}$ & $0.28_{-0.17}^{+0.16}$ & $52.500934_{-0.000049}^{+0.000048}$ & $0.287_{-0.033}^{+0.036}$ & $0.744_{-0.083}^{+0.095}$ \\
1314.01 & $0.01180_{-0.00031}^{+0.00068}$ & $1.965_{-0.291}^{+0.093}$ & $0.38_{-0.29}^{+0.30}$ & $8.575116_{-0.000031}^{+0.000030}$ & $0.54_{-0.20}^{+0.26}$ & $0.32_{-0.19}^{+0.28}$ \\
1894.01 & $0.01739_{-0.00039}^{+0.00087}$ & $1.79_{-0.23}^{+0.10}$ & $0.41_{-0.31}^{+0.25}$ & $5.2879067_{-0.0000085}^{+0.0000085}$ & $0.361_{-0.092}^{+0.115}$ & $0.51_{-0.15}^{+0.21}$ \\
2133.01 & $0.019429_{-0.000066}^{+0.000109}$ & $1.6382_{-0.0348}^{+0.0066}$ & $0.093_{-0.093}^{+0.160}$ & $ 6.2467332_{-0.0000046}^{+0.0000046}$ & $0.977_{-0.044}^{+0.022}$ & $0.044_{-0.025}^{+0.019}$  \\
2481.01 & $0.01460_{-0.00050}^{+0.00092}$ & $1.70_{-0.31}^{+0.11}$ & $0.39_{-0.27}^{+0.30}$ & $33.84513_{-0.00052}^{+0.00086}$ & $0.24_{-0.11}^{+0.25}$ & $0.85_{-0.24}^{+0.11}$ \\
2640.01 & $0.01619_{-0.00043}^{+0.00064}$ & $2.33_{-0.43}^{+0.16}$ & $0.49_{-0.37}^{+0.29}$ & $33.17354_{-0.00014}^{+0.00014}$ & $0.104_{-0.072}^{+0.118}$ & $0.52_{-0.33}^{+0.33}$ \\
\hline\hline 
\end{tabular}
\label{tab:fittedtable} 
\end{table*}

%% file: rhotable.tex
\begin{table*}
\caption{\emph{Asterodensity profiling parameters for the 41 KOIs in our sample.
\rhoA\ values come from \citet{huber:2013}. Columns 4, 5 and 6 are computed
using Equations~39 \& 17 of \citet{AP:2014} and Equation~\ref{eqn:rhoalt} of
this work, respectively. Classification denotes AP which can explain the
observations, where ``N'' denotes no AP effect required. KOIs with a $^*$ imply
that $b>(1-p)$ in more than half the posteriors samples, meaning the AP
equations become invalid. Rows above the horizontal line have $\log g>3.7$ and
those below have $\log g \leq 3.7$.}} 
\centering 
\begin{tabular}{l l l l l l l l l l} 
\hline\hline 
KOI & \vline & \rhoO\,[kg\,m$^{-3}$] & \rhoA\,[kg\,m$^{-3}$] & \vline & $e_{\mathrm{min}}$ & $B$ & $\rho_{\star,\mathrm{alt,max}}$\,[kg\,m$^{-3}$] & \vline & Classification \\ [0.5ex] 
\hline
1.01 & \vline & $1532_{-34}^{+48}$ & $1530\pm30$ & \vline & $0.0074_{-0.0051}^{+0.0081}$ & $0.991_{-0.087}^{+0.086}$ & $2636_{-55}^{+72}$ & \vline & N \\
2.01 & \vline & $276.47_{-0.37}^{+0.37}$ & $271.2\pm3.2$ & \vline & $0.0064_{-0.0038}^{+0.0040}$ & $0.969_{-0.019}^{+0.019}$ & $1539.1_{-3.8}^{+3.7}$ & \vline & N \\
7.01 & \vline & $392_{-43}^{+53}$ & $427\pm13$ & \vline & $0.034_{-0.023}^{+0.036}$ & $1.13_{-0.19}^{+0.21}$ & $11600_{-2200}^{+3000}$ & \vline & N \\
64.01 & \vline  & $290_{-14}^{+15}$ & $123.8\pm3.3$ & \vline & $0.277_{-0.017}^{+0.018}$ & $0.192_{-0.016}^{+0.017}$ & $753_{-55}^{+58}$ & \vline & PE/FP \\
69.01 & \vline & $1440_{-160}^{+180}$ & $1640\pm10$ & \vline & $0.042_{-0.030}^{+0.039}$ & $1.19_{-0.18}^{+0.21}$ & $89000_{-17000}^{+22000}$ & \vline & N \\
75.01 & \vline & $71.4_{-1.6}^{+1.7}$ & $115.2\pm4.0$ & \vline & $0.158_{-0.014}^{+0.014}$ & $2.13_{-0.15}^{+0.16}$ & $678_{-28}^{+31}$ & \vline & FP/PE \\
87.01 & \vline & $2900_{-1900}^{+3600}$ & $1458\pm30$ & \vline & $0.26_{-0.19}^{+0.20}$ & $0.37_{-0.24}^{+1.42}$ & $66000_{-56000}^{+230000}$ & \vline & N \\
97.01 & \vline & $233.8_{-1.8}^{+1.8}$ & $245\pm15$ & \vline & $0.018_{-0.013}^{+0.018}$ & $1.08_{-0.11}^{+0.12}$ & $1119_{-15}^{+15}$ & \vline & N \\
98.01 & \vline & $168_{-3.1}^{+3.2}$ & $224\pm14$ & \vline & $0.096_{-0.022}^{+0.021}$ & $1.56_{-0.16}^{+0.17}$ & $1614_{-53}^{+56}$ & \vline & PB/PE/FP \\
107.01 & \vline & $478_{-136}^{+75}$ & $427\pm32$ & \vline & $0.065_{-0.043}^{+0.048}$ & $0.86_{-0.17}^{+0.52}$ & $19900_{-9100}^{+6200}$ & \vline & N \\
113.01* & \vline & $5050_{-720}^{+580}$ & $382\pm24$ & \vline & $0.696_{-0.028}^{+0.021}$ & - & - & \vline & - \\
118.01 & \vline & $920_{-390}^{+180}$ & $581\pm30$ & \vline & $0.170_{-0.109}^{+0.048}$ & $0.53_{-0.12}^{+0.60}$ & $55000_{-35000}^{+21000}$ & \vline & PE/FP \\
122.01 & \vline & $1760_{-177}^{+48}$ & $540\pm19$ & \vline & $0.372_{-0.029}^{+0.015}$ & $0.201_{-0.014}^{+0.029}$ & $77700_{-13500}^{+3700}$ & \vline & PE/FP \\
257.01 & \vline & $966_{-67}^{+76}$ & $990\pm34$ & \vline & $0.019_{-0.013}^{+0.021}$ & $1.04_{-0.13}^{+0.14}$ & $8500_{-1000}^{+1300}$ & \vline & N \\
263.01 & \vline & $1070_{-520}^{+1400}$ & $378\pm11$ & \vline & $0.33_{-0.21}^{+0.22}$ & $0.24_{-0.16}^{+0.35}$ & $37000_{-26000}^{+147000}$ & \vline & PE/FP \\
268.01 & \vline & $491_{-117}^{+49}$ & $662\pm21$ & \vline & $0.100_{-0.033}^{+0.088}$ & $1.53_{-0.20}^{+0.73}$ & $20500_{-7900}^{+3900}$ & \vline & PB/PE/FP \\
269.01 & \vline & $113_{-44}^{+79}$ & $605\pm18$ & \vline & $0.51_{-0.14}^{+0.11}$ & $12.1_{-7.2}^{+20.9}$ & $3700_{-2200}^{+6300}$ & \vline & FP/PE \\
273.01 & \vline & $910_{-150}^{+190}$ & $1193\pm25$ & \vline & $0.091_{-0.056}^{+0.061}$ & $1.65_{-0.50}^{+0.81}$ & $4600_{-1200}^{+1700}$ & \vline & N \\
276.01 & \vline & $2010_{-790}^{+1440}$ & $898\pm32$ & \vline & $0.26_{-0.16}^{+0.16}$ & $0.32_{-0.16}^{+0.32}$ & $61000_{-37000}^{+104000}$ & \vline & PE/FP \\
280.01 & \vline & $1090_{-140}^{+200}$ & $1281\pm27$ & \vline & $0.059_{-0.039}^{+0.045}$ & $1.29_{-0.30}^{+0.33}$ & $12200_{-2700}^{+4400}$ & \vline & PB/PE/FP \\
281.01 & \vline & $160_{-57}^{+124}$ & $459\pm15$ & \vline & $0.34_{-0.18}^{+0.12}$ & $4.6_{-2.6}^{+5.1}$ & $5600_{-3100}^{+10600}$ & \vline & FP/PE \\
288.01 & \vline & $249_{-69}^{+92}$ & $221.2\pm2.7$ & \vline & $0.081_{-0.057}^{+0.068}$ & $0.85_{-0.29}^{+0.49}$ & $15100_{-6900}^{+11900}$ & \vline & N \\
319.01 & \vline & $221_{-13}^{+13}$ & $206\pm15$ & \vline & $0.027_{-0.019}^{+0.028}$ & $0.87_{-0.15}^{+0.18}$ & $637_{-45}^{+50}$ & \vline & N \\
975.01 & \vline & $564.9_{-7.9}^{+5.8}$ & $288.6\pm8.7$ & \vline & $0.220_{-0.010}^{+0.010}$ & $0.405_{-0.017}^{+0.018}$ & $105800_{-2600}^{+1900}$ & \vline & PE/FP \\
1282.01 & \vline & $432_{-97}^{+39}$ & $392\pm21$ & \vline & $0.048_{-0.031}^{+0.034}$ & $0.88_{-0.13}^{+0.40}$ & $5750_{-2110}^{+860}$ & \vline & N \\
1537.01 & \vline & $288_{-169}^{+92}$ & $314\pm11$ & \vline & $0.078_{-0.053}^{+0.235}$ & $1.12_{-0.35}^{+2.67}$ & $49000_{-41000}^{+33000}$ & \vline & N \\
1618.01 & \vline & $542_{-155}^{+60}$ & $524\pm13$ & \vline & $0.040_{-0.027}^{+0.061}$ & $0.96_{-0.13}^{+0.55}$ & $179000_{-86000}^{+29000}$ & \vline & N \\
1621.01 & \vline & $1000_{-530}^{+200}$ & $244\pm14$ & \vline & $0.436_{-0.182}^{+0.051}$ & $0.150_{-0.034}^{+0.251}$ & $86000_{-65000}^{+34000}$ & \vline & PE/FP \\
1890.01 & \vline & $478_{-218}^{+66}$ & $450\pm17$ & \vline & $0.060_{-0.035}^{0.121}$ & $0.93_{-0.16}^{+1.21}$ & $60000_{-41000}^{+16000}$ & \vline & N \\
1924.01 & \vline & $65.3_{-2.5}^{+2.2}$ & $131.9\pm4.3$ & \vline & $0.230_{-0.015}^{+0.016}$ & $2.58_{-0.16}^{+0.18}$ & $25900_{-2200}^{+2000}$ & \vline & FP/PE \\
1962.01 & \vline & $11600_{-4600}^{+1900}$ & $477\pm45$ & \vline & $0.785_{-0.073}^{+0.023}$ & $0.0126_{-0.0028}^{+0.0106}$ & $207000_{-124000}^{+59000}$ & \vline & PE/FP \\
\hline
371.01* & \vline & $710_{-170}^{+270}$ & $66.1\pm1.6$ & \vline & $0.658_{-0.054}^{+0.057}$ & - & - & \vline & - \\
674.01 & \vline & $187.8_{-15.7}^{+5.6}$ & $84.0\pm3.7$ & \vline & $0.259_{-0.027}^{+0.019}$ & $0.325_{-0.026}^{+0.041}$ & $3510_{-520}^{+190}$ & \vline & PE/FP \\
1222.01 & \vline & $540_{-340}^{+270}$ & $38.5\pm1.9$ & \vline & $0.707_{-0.211}^{+0.063}$ & $0.029_{-0.012}^{+0.070}$ & - & \vline & FP \\
1230.01 & \vline & $99.9_{-1.9}^{+2.0}$ & $7.091\pm0.067$ & \vline & $0.7073_{-0.0035}^{+0.0036}$ & $0.02216_{-0.00057}^{+0.00057}$ & $646_{-22}^{+23}$ & \vline & PE/FP \\
1299.01 & \vline & $131_{-23}^{+14}$ & $26.50\pm0.49$ & \vline & $0.488_{-0.051}^{+0.025}$ & $0.109_{-0.013}^{+0.032}$ & $3410_{-1000}^{+700}$ & \vline & PE/FP \\
1314.01 & \vline & $92_{-45}^{+22}$ & $42.33\pm0.82$ & \vline & $0.259_{-0.154}^{+0.062}$ & $0.345_{-0.086}^{+0.510}$ & $7900_{-5700}^{+3900}$ & \vline & N \\
1894.01 & \vline & $62_{-26}^{+16}$ & $36.5\pm1.3$ & \vline & $0.180_{-0.123}^{+0.069}$ & $0.49_{-0.13}^{+0.54}$ & $2900_{-1900}^{+1700}$ & \vline & PE/FP \\
2133.01 & \vline & $43.47_{-3.35}^{+0.67}$ & $6.81\pm0.32$ & \vline & $0.547_{-0.019}^{+0.013}$ & $0.0813_{-0.0061}^{+0.0092}$ & $2113_{-297}^{+49}$ & \vline & FP \\
2481.01 & \vline & $50_{-26}^{+14}$ & $1.999\pm0.096$ & \vline & $0.791_{-0.104}^{+0.030}$ & $0.0128_{-0.0037}^{0.018}$ & $3200_{-2400}^{+1600}$ & \vline & FP  \\
2640.01 & \vline & $212_{-133}^{+97}$ & $4.29\pm0.10$ & \vline & $0.862_{-0.113}^{+0.029}$ & $0.0051_{-0.0020}^{+0.0127}$ & $10200_{-8400}^{+9800}$ & \vline & FP \\
\hline\hline 
\end{tabular}
\label{tab:rhotable} 
\end{table*}